\documentclass[twocolumn,showpacs,preprintnumbers]{revtex4}
\usepackage{amssymb}
\usepackage{amsmath}
\usepackage{graphicx}
\usepackage{dcolumn}
\usepackage{bm}
\usepackage{color}
\usepackage{soul}

\soulregister\cite7
\soulregister\ref7

\begin{document}

\title{Quantum tricritical point emerging in the spin-boson model with two
dissipative spins in staggered biases}
\author{Yan-Zhi Wang$^{1, 2}$, Shu He$^{3}$, Liwei Duan$^{1,4}$, and Qing-Hu
Chen$^{1,5,*}$}

\address{
$^{1}$ Zhejiang Province Key Laboratory of Quantum Technology and Device, Department of Physics, Zhejiang University, Hangzhou 310027, China \\
$^{2}$ School of Physics and Electronic information, Anhui  Normal University, Wuhu  241002, China\\
$^{3}$ Department of Physics and Electronic Engineering, Sichuan Normal University, Chengdu 610066, China\\
$^{4}$ Department of Physics, Zhejiang Normal University, Jinhua 321004,  China \\
$^{5}$  Collaborative Innovation Center of Advanced Microstructures, Nanjing University, Nanjing 210093, China
 }\date{\today }

\begin{abstract}
We study the spin-boson model (SBM) with two spins in staggered biases by a
numerically exact method based on variational matrix product states. Several
observables such as the magnetization, the entanglement entropy between the
two spins and the bosonic environment, the ground-state energy, as well as
the correlation function for two spins are calculated exactly. The
characteristics of these observables suggest that the staggered biases can
drive the 2nd-order quantum phase transition (QPT) to the 1st-order QPT in
the sub-Ohmic SBM, while the Kosterlitz-Thouless QPT in the Ohmic SBM
goes directly to the 1st-order one. A quantum tricritical point, where the
continuous QPT meets the 1st-order one, can  then be  detected. It is found
that the staggered biases would not change the universality of {
the phase transition in this model} below the quantum tricritical point.
\end{abstract}

\maketitle

\section{Introduction}

In the light-matter interacting systems, it is well known for a long time
that the prototype Dicke model \cite{Dicke} and the spin-boson model (SBM)
\cite{Leggett,Breuer,weiss1} can display quantum phase transitions (QPTs) at
strong coupling between the two-level systems (qubits) and the cavity or the
bosonic baths. Recently, it has even been proposed that the quantum Rabi
model only consisting of one qubit and a single-mode cavity can undergo a
QPT in the infinite ratio of the qubit and cavity frequencies~\cite{QRM},
which further inspires a surge of studies for the so-called finite-component
QPT \cite{Liu,zhengsb,Zu-Jian,Felicetti,yfxie}. It is generally accepted
that the Dicke model and the quantum Rabi model only experience a single QPT
from the normal to the superradiant phase with the same critical behavior,
and the SBM exhibits the single QPT from the delocalized to the localized
phases with the spectra function dependent critical exponents ~\cite{Bulla}.

Theoretically, to obtain a rich phase diagram of quantum phases in the
light-matter interacting systems, one can generalize these prototype models
to their variants. Generally, the QPT only appears in the Dicke model in the
thermodynamic limit, i.e. the qubit number $N\rightarrow \infty $,
exhibiting the mean-field critical behavior. The generalized Dicke models,
such as the anisotropic Dicke model ~\cite{yejw,ciute,Peng}, the anisotropic
Dicke model with the nonlinear Stark coupling terms~ \cite{zhiqiang}, and
the Dicke model where infinite atoms are separated equally into two parts
each experiencing the opposite equal biases ~\cite{puhan} have been recently
studied by several groups. In these generalized Dicke models, both the 1st-
and 2nd-order QPTs are observed. More recently, the existence of the
finite-component multicriticality is demonstrated in a generalized Dicke
model with a finite number of atoms at an extremely large detunings ~\cite%
{GFZhang}.

A quantum tricritical point (QuTP)~\cite{Griffiths} is seldomly supported in
the solid-state materials, and is almost impossible to appear in the
prototype models of the light-matter interacting systems. Interestingly, it
has been found to exist in anisotropic Dicke model ~\cite{ciute} and the
isotropic Dicke model with staggered biases \cite{puhan}~. In the former
model, the QuTP lies at the symmetric line of the superradiant
\textquotedblleft electric\textquotedblright\ and \textquotedblleft
magnetic\textquotedblright\ phases, which can be mapped mutually by
interchanging the rotating-wave term and the counterrotating one, while in
the latter model, it was demonstrated that the field can drive the 2nd-order
QPT  to the 1st-order one, thus the 2nd-order critical line can meet the
1st-order critical line at the QuTP \cite{puhan}.

In the SBM with single qubit, the 2nd-order QPT from the delocalized phase,
where spin has  equal probability in the two states, to the localized
phase, in which the spin prefers to stay in one of the two states, has been
studied extensively ~\cite{Bulla,QMC,ED,Vojta1,Zhang,Chin,guo2012critical,Frenzel,CRduan,HeShu1,zhou,Shao,Fazio}
. Recently, the anisotropic sub-Ohmic SBM has also been studied by the
present authors \cite{ASBM}. It is generally accepted that the continuous
QPT with mean-field exponents is found for the power of the bath spectral
function $s<1/2$ ~\cite{QMC,ED,Zhang}, with nontrivial exponents for $1/2<s<1
$~\cite{Bulla,Vojta1}. The Kosterlitz-Thouless (KT) phase transition occurs
for $s=1$ \cite{Leggett}, and no phase transition happens for $s>1$.

The SBM
has been generalized by increasing the number of spins, such as the SBM with
two spins \cite{Lehur2,lvzheng}, and a finite number of spins even in the
limit $N\rightarrow \infty $ \cite{Regir2014}. It has been found that the
critical behavior of QPT is not changed with the increasing number of spins.
Only in the limit $N\rightarrow \infty $, the universality class of the
transition changes into mean-field behavior.

We will study the criticality of the generalized SBM with two spins in
staggered biases. Our goals are two-fold. { Since the staggered
biases result in the QuTP in the generalized Dicke model~\cite{puhan} and
even multicriticality in a finite number of qubits collectively coupled to a
single mode cavity at an extremely large detuning \cite{GFZhang}, we first
explore whether the QuTP can emerge in the two-qubit SBM with staggered
biases. In the original SBM, the continuous QPT occurs in the sub-Ohmic
baths \cite{QMC,ED,Zhang,Bulla,Vojta1,guo2012critical} where the critical
exponents are bath dependent, while the KT phase transition in the Ohmic
bath~\cite{Leggett,lehur_kt}. This picture is not changed for the SBM with a
finite number of spins without biases \cite{Lehur2,lvzheng,Regir2014}. We
then examine \ whether the presence of the staggered biases would change the
universality class of these continuous QPTs in the SBM with two spins.}

In this paper, we will extend the variational matrix product state (VMPS)
approach ~\cite{guo2012critical} to study the two-spin-boson model (2SBM)
with both sub-Ohmic and Ohmic baths. The paper is organized as follows. In
Sec. II, we introduce the 2SBM in the staggered biases for two spins along
the opposite directions and the VMPS approach briefly. In Sec. III, we study
the QPTs of the 2SBM in both sub-Ohmic and Ohmic baths with the staggered
biases. For the sub-Ohmic bath, we choose two typical powers of the spectra
function of the bath, which are, respectively, corresponding to the
mean-field and interacting  critical nature of the QPTs in the single SBM.
The order parameter and the entanglement entropy between the two qubits and
the bosonic bath are extensively calculated. The critical exponents for the
order parameter are also analyzed. A QuTP separated by the 2nd-order (KT
type) critical lines and the 1st-order ones for the sub-Ohmic (Ohmic) baths
are revealed by several independent evidences from different observables.
Finally, we present a brief summary in the last section.

\section{2SBM with staggered biases and methodologies}

The 2SBM Hamiltonian can be written as (the reduced Planck constant is set $%
\hbar =1$)
\begin{eqnarray}
\hat{H} &=&\sum_{i=1,2}\frac{1}{2}\left( \Delta \sigma _{i}^{z}-\left(
-1\right) ^{i}\epsilon \sigma _{i}^{x}\right) +\sum_{k}\omega
_{k}a_{k}^{\dag }a_{k}  \notag \\
&&+\frac{1}{2}\sum_{k}g_{k}\left( a_{k}^{\dag }+a_{k}\right) \left( \sigma
_{1}^{x}+\sigma _{2}^{x}\right),  \label{Hamiltonian}
\end{eqnarray}%
where $\sigma _{i=1,2}^{j}$ ($j=x,y,z$) are the Pauli matrices for spins $1$
and $2$, $\Delta $ is the qubit frequency, $\left( -1\right) ^{i}\epsilon $
before $\sigma _{i}^{x}$ represents the staggered biases along the x axis
for the two spins, $a_{k}$ ($a_{k}^{\dag }$) is the bosonic annihilation
(creation) operator which can annihilate (create) a boson with frequency $%
\omega _{k}$, and $g_{k}$ denotes the coupling strength between the qubit
and the bosonic bath, which is usually characterized by the power-law
spectral density $J(\omega )$,
\begin{equation}
J(\omega )=\pi \sum_{k}g_{k}^{2}\delta (\omega -\omega _{k})=2\pi \alpha
\omega _{c}^{1-s}\omega ^{s}\Theta (\omega _{c}-\omega ),
\end{equation}%
where $\alpha $ is a dimensionless coupling constant, $\omega _{c}$ is the
cutoff frequency, and $\Theta (\omega _{c}-\omega )$ is the Heaviside step
function. The power of the spectral function $s$ classifies the reservoir
into super-Ohmic $\left( s>1\right) $, Ohmic $\left( s=1\right) $, and
sub-Ohmic $\left( s<1\right) $ types. This model is illustrated in Fig. \ref%
{view} where the x axis is in a horizontal line.

\begin{figure}[tph]
\centerline{\includegraphics[scale=0.4]{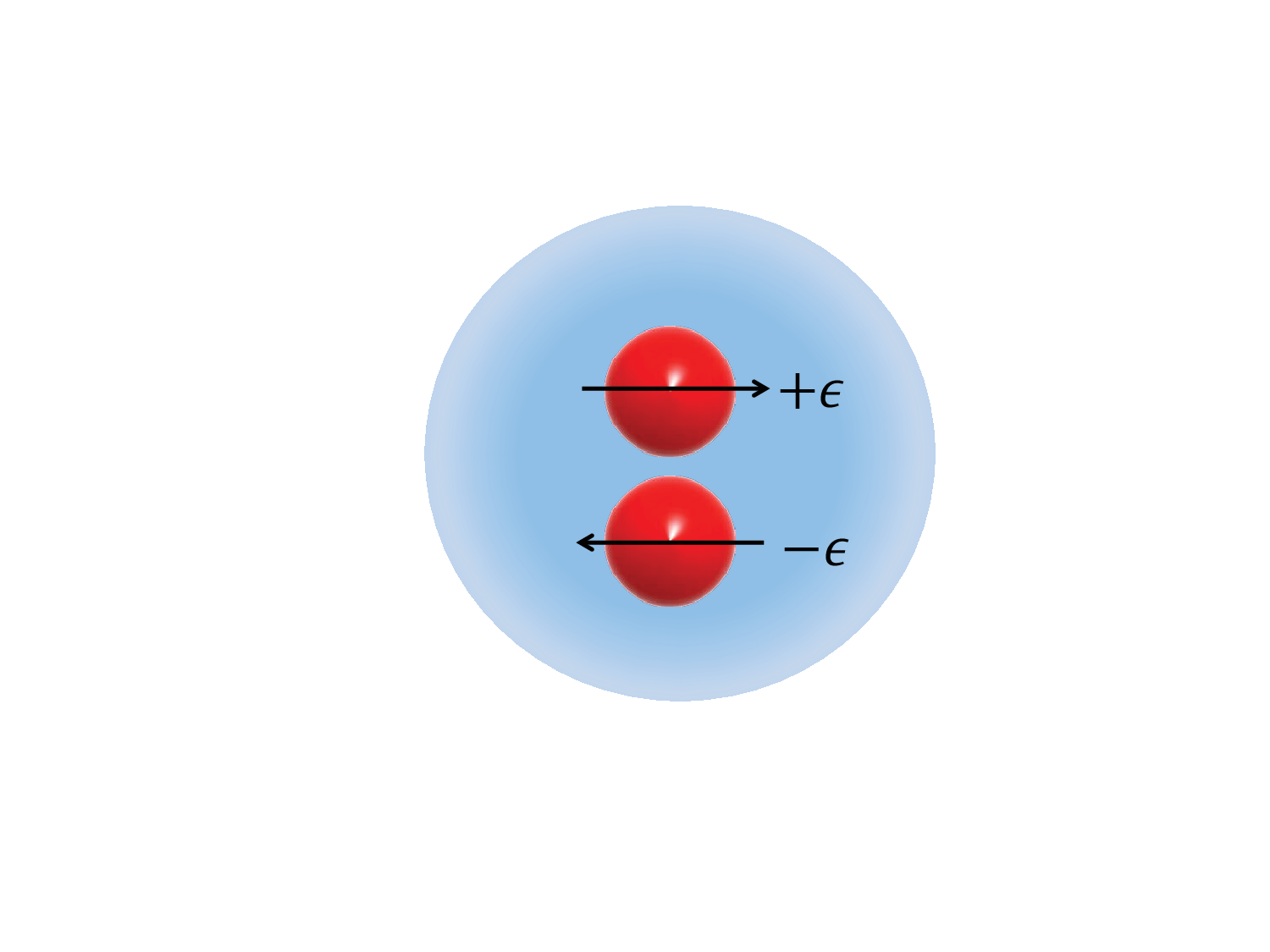}}
\caption{ (Color online) Illustration of the two-spin-boson model with
staggered biases $\pm\protect\epsilon$ along the x direction. The two spins
denoted by red spheres interact with a common continuous bosonic reservoir
represented by the big blue region. No direct interaction between spins is
considered. }
\label{view}
\end{figure}

The introduced staggered biases to the two spins do not break the parity ($%
Z_{2}$) symmetry in the 2SBM. The parity operator is defined as%
\begin{equation}
\hat{\Pi}=\left[ \frac{\sigma _{1}^{z}\sigma _{2}^{z}+1}{2}-\left( \sigma
_{1}^{+}\sigma _{2}^{-}+\sigma _{1}^{-}\sigma _{2}^{+}\right) \right] \exp
\left( i\pi \sum_{k}a_{k}^{\dag }a_{k}\right) ,  \label{parity}
\end{equation}%
where $\sigma _{i=1,2}^{\pm }=\left( \sigma _{i}^{x}\pm i\sigma
_{i}^{y}\right) /2$. Note that in the presence of the staggered biases, the
parity operator is more complicated than that for $\epsilon =0$, $\hat{\Pi}%
_{\epsilon =0}=\exp \left[ i\pi \left( \sum_{k}a_{k}^{\dag }a_{k}+\left(
\sigma _{1}^{z}+\sigma _{2}^{z}\right) /2+1\right) \right] $, due to the
absence of the collective spin. The parity operator $\hat{\Pi}$ has two
eigenvalues \ $\pm 1$, corresponding to the even and odd parity in the
symmetry conserved phases. The average value of the parity may become zero
due to the quantum fluctuations in the symmetry broken phase.

To apply VMPS in the 2SBM in staggered biases, the logarithmic
discretization of the spectral density of the continuum bath~\cite{Bulla}
with discretization parameter $\Lambda >1$ is performed first, followed by
using orthogonal polynomials as described in Ref.~\cite{Plenio_Chin}, the
2SBM can be mapped into the representation of an 1D semi-infinite chain with
nearest-neighbor interaction ~\cite{Friend}. Thus, Hamiltonian (\ref%
{Hamiltonian}) can be written as:
\begin{eqnarray}
H_{\text{chain}} &=&\frac{\Delta }{2}\left( \sigma _{1}^{z}+\sigma
_{2}^{z}\right) +\frac{\epsilon }{2}\left( \sigma _{1}^{x}-\sigma
_{2}^{x}\right)  \notag \\
&&+\frac{c_{0}}{2}(b_{0}+b_{0}^{\dag })\left( \sigma _{1}^{x}+\sigma
_{2}^{x}\right)  \notag \\
&&+\sum_{n=0}^{L-2}[\varepsilon _{n}b_{n}^{\dag }b_{n}+t_{n}(b_{n}^{\dag
}b_{n+1}+b_{n+1}^{\dag }b_{n})],  \label{Hamitrans}
\end{eqnarray}%
where $b_{n}^{\dag }$($b_{n}$) is the creation (annihilation) operator for a
new set of boson modes in a transformed representation with $\varepsilon
_{n} $ describing the frequency on chain site $n$, $t_{n}$ the
nearest-neighbor hopping parameter, and $c_{0}$ the effective coupling
strength between the spin and the new effective bath. For more details, one
may refer to Ref.~\cite{Plenio_Chin}.

Then as introduced in Refs. ~\cite{VMPS1,VMPS2}, we employ the standard matrix
product representation with the optimized boson basis $\left\vert \widetilde{%
n}_{k}\right\rangle$ through an additional isometric map with truncation
number $d_{opt}\ll d_{n}$ like in Refs.~\cite{guo2012critical,Friend} to
study the quantum criticality of 2SBM. Each site in the 1D chain can be
described by the matrix $M_s$, which is optimized through sweeping the 1D
chain iteratively to obtain the ground state, and $D_{n}$ is the bond
dimension for matrix $M_s$ with the open boundary condition, bounding the
maximal entanglement in each subspace.

For the data presented below, we typically choose the same model parameters
in Refs. \cite{guo2012critical,wangyz,ASBM}, as $\Delta =0.1$, $\omega _{c}=1$%
, the logarithmic discretization parameter $\Lambda =2$, the length of the
semi-infinite chain $L=50$, and optimized truncation numbers $d_{opt}=12$.
In addition, we adjust the bond dimension as $D_{max}=20,40,$ and $20$ for $%
s=0.3,0.7$, and $\ 1$, respectively, which are sufficient to obtain the
converged results for the problems concerned. {\ Actually, the convergence
thresholds for the bond dimensions are $D=12$ for $s=0.3$, and $D=20$ for $%
s=0.7$ and $1$ for the all observables. Only in the critical regime, we use
larger bond dimensions $D=20$ for $s=0.3$ and $D=40$ for $s=0.7$ to improve
computational accuracy where the relative error is less than $10^{-7}$ for
the energy and $10^{-5}$ for the magnetization, so that we can evaluate the
critical exponents precisely. The evidence for the full convergence of our
VMPS results here is similar to that demonstrated in the Appendix of Ref. ~%
\cite{wangyz}, and will not be repeated in this paper }

The information of the ground-state can also be  described by the von Neumann
{entropy $S_{E}$ of the 2SBM, which characterizes the entanglement between
two spins and the bosonic bath }%
\begin{equation}
{{S_{E}}=-Tr\left( {{\rho _{spin}\log \rho _{spin}}}\right) },
\label{entropy}
\end{equation}%
{where $\rho _{spin}$ is the reduced density matrix for the two spins. }

The averaged total magnetization
\begin{equation}
M= \left( \left\langle \sigma _{1}^{x}\right\rangle +\left\langle \sigma
_{2}^{x}\right\rangle \right) /2,  \label{order}
\end{equation}%
can be regarded\ as the order parameter, which can be used to characterize
the essential nature of the 2nd-order QPTs. However, $M$ is hardly employed
to distinguish the KT and the 1st-order QPTs, because it would suddenly drop
to zero in both cases.

\section{Results and discussions}

\subsection{Sub-Ohmic bath ($s<1)$}

\begin{figure}[tph]
\centerline{\includegraphics[scale=0.4]{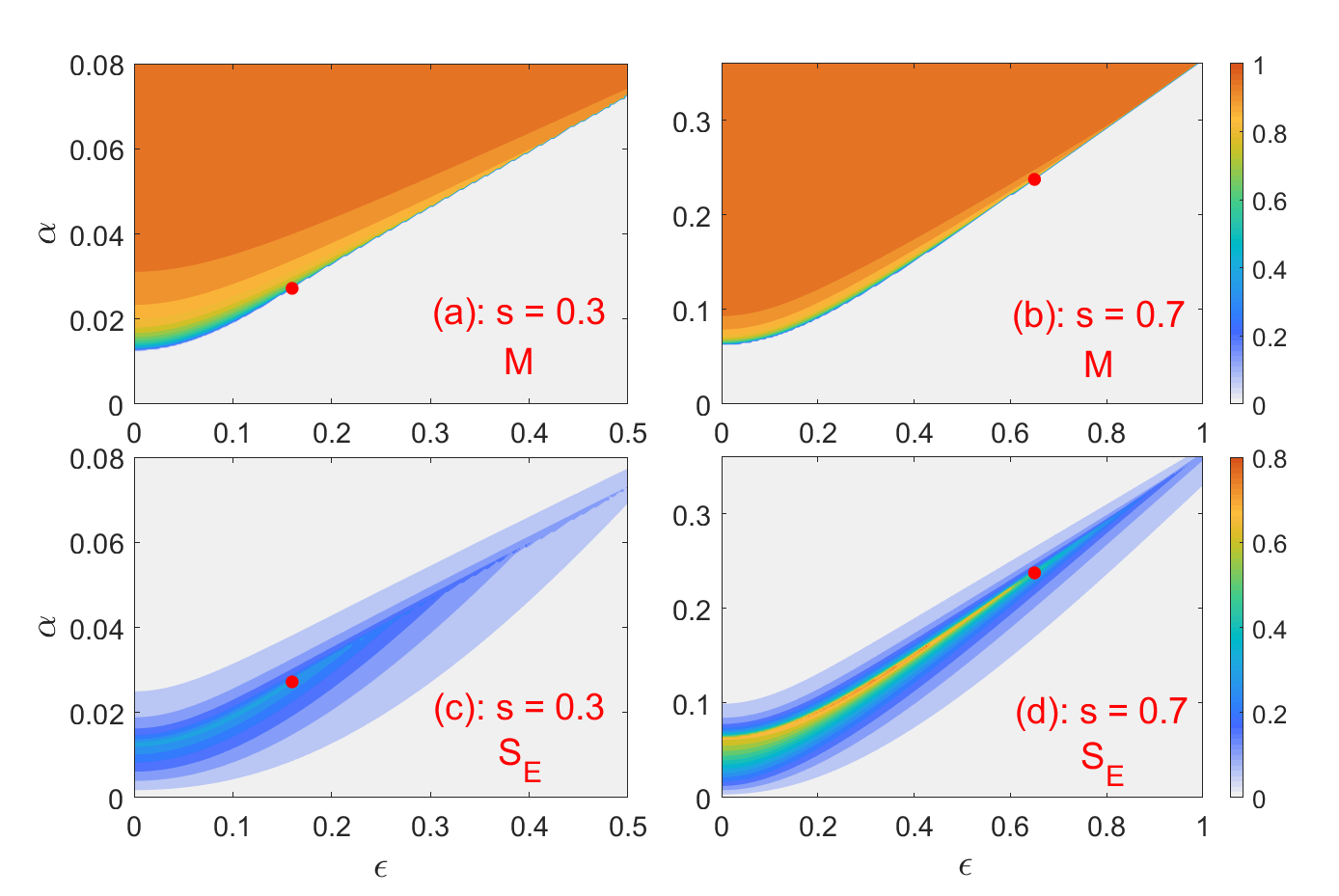}}
\caption{ (Color online) (Upper panels) Phase diagram in the $\protect%
\epsilon -\protect\alpha $ plane for 2SBM drawn from the Magnetization $M$:
delocalized phases ($M=0$) and the localized phase ($M\neq 0$). (Lower
panels) Entanglement entropy $S_{E}$. The power of the spectral function is
(left) $s=0.3$ and (right) $0.7$. $\Delta =0.1$, $\protect\omega _{c}=1$.
The QuTP is marked by a red dot, which separates the intersection of the
2nd- and 1st-order phase transition. The parameters used in the VMPS
approach are $\Lambda =2$, $L=50$, $d_{opt}=12$, and $D=20$ for $s=0.3,0.7$.
}
\label{phase_diagram}
\end{figure}

The single SBM expects a mean-field critical behavior for $s<1/2$, and a
nonclassical one for $s>1/2$, so we focus on two typical powers of the
spectral function $s=0.7$ and $0.3$ for the sub-Ohmic case. The entire
critical lines can be mapped out by the onset of the nonzero order parameter
$M=\left( \left\langle \sigma _{1}^{x}\right\rangle +\left\langle \sigma
_{2}^{x}\right\rangle \right) /2$. By this criterion, the phase diagrams of
2SBM with staggered biases are summarized in the $\epsilon -\alpha $ plane
in the upper panels of Fig. \ref{phase_diagram} for $s=0.3$ (left) and $0.7$
(right), respectively. To confirm the phase diagram more convincingly, we
also display the entanglement entropy $S_{E}$ between the two spins and the
bath,  an alternative widely used tool in the location of QPTs, in the lower
panels of Fig. \ref{phase_diagram}. The entropy displays a sharp
nonanalyticity at the phase transition ~\cite{Chakravarty,lehur,Chin}. The
ridge line of $S_{E}$ obviously shows a sharp nonanalyticity, which exactly
coincides with the critical line obtained by the order parameter. {\ At
either infinite coupling strength or infinite bias, the entanglement becomes
zero due to the decoupling of systems and environments in two extreme cases.}
\begin{figure}[tph]
\centerline{\includegraphics[scale=0.4]{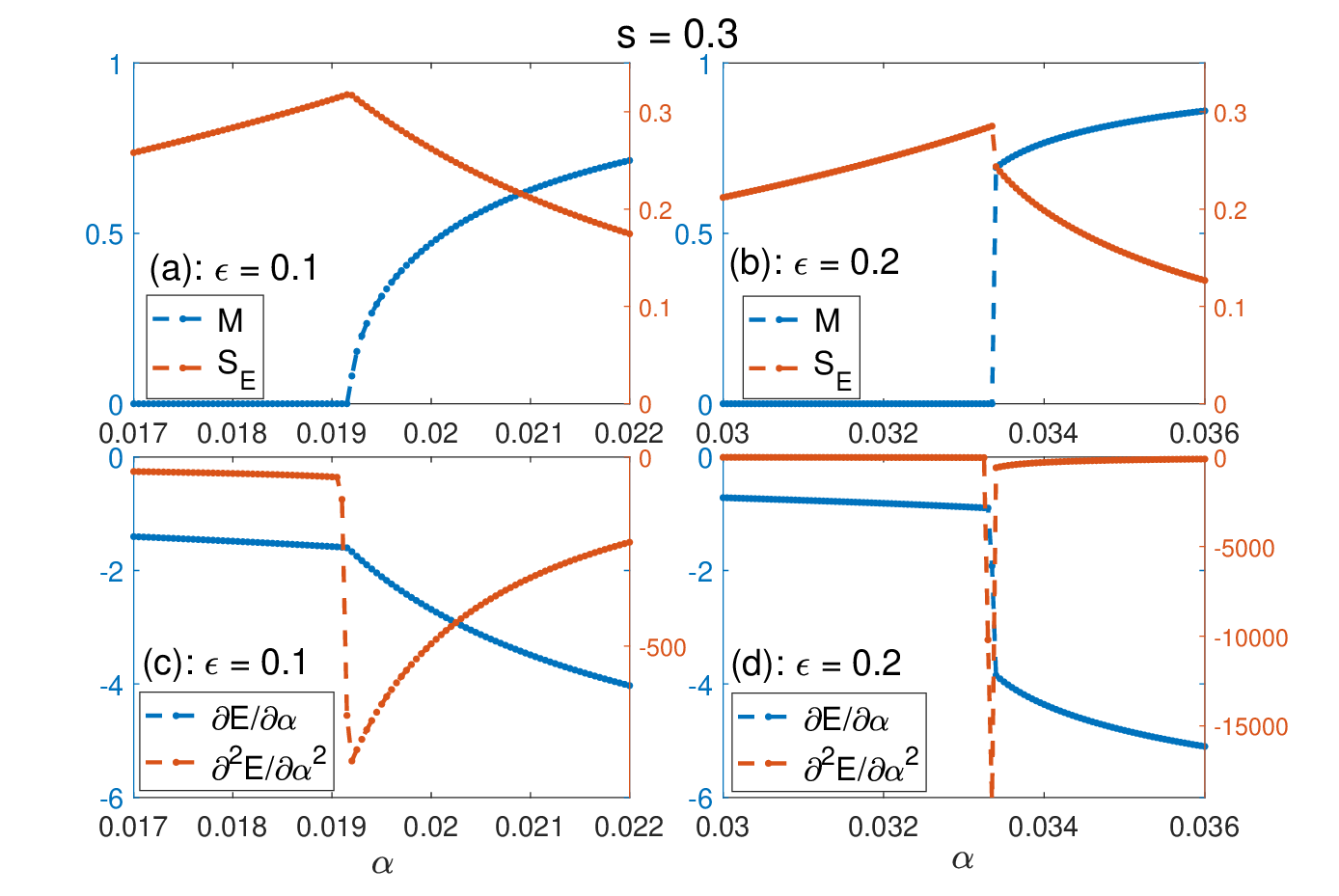}}
\caption{ (Color online) Magnetization $M$, entanglement entropy $S_{E}$
(upper panels), the first- and second-order derivatives of the ground-state
energy (lower panels) as a function of $\protect\alpha $ for $\protect%
\epsilon =0.1$ (left) and $\protect\epsilon =0.2$ (right) {\ for the
sub-Ohmic bath with $s=0.3$} by VMPS approach. $\Delta =0.1$, $\protect%
\omega _{c}=1$, $\Lambda =2$, $L=50$, $d_{opt}=12$, and $D=20$. }
\label{QPT_s03}
\end{figure}

\begin{figure}[tph]
\centerline{\includegraphics[scale=0.4]{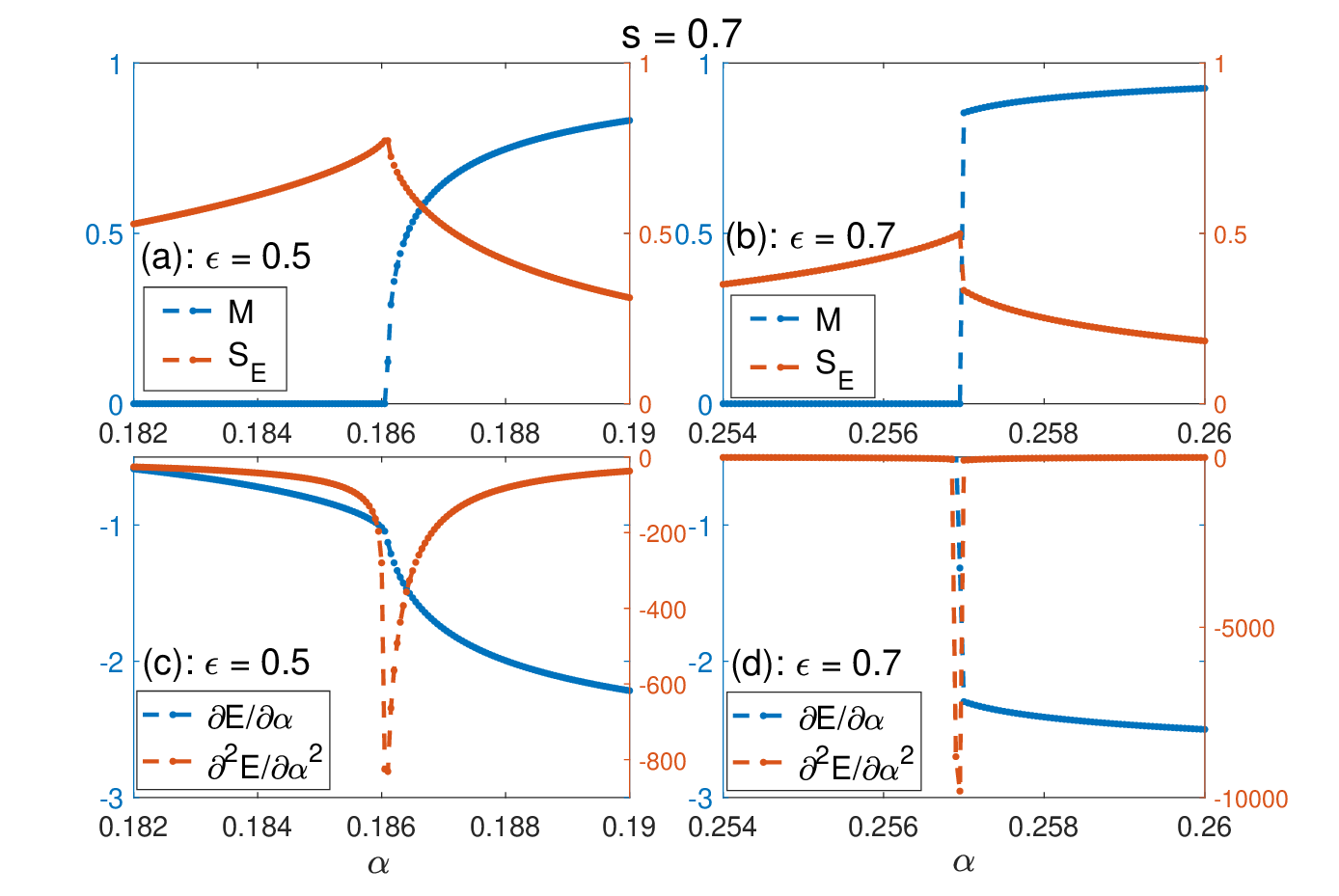}}
\caption{ (Color online) Magnetization $M$, entanglement entropy $S_{E} $
(upper panels), the first- and second-order derivatives of the ground-state
energy (lower panels) as a function of $\protect\alpha $ for $\protect%
\epsilon =0.5$ (left) and $\protect\epsilon =0.7$ (right) {\ for the
sub-Ohmic bath with $s=0.7$ } by VMPS approach. $\Delta =0.1$, $\protect%
\omega _{c}=1$, $\Lambda =2$, $L=50$, $d_{opt}=12$, and $D=20$. }
\label{QPT_s07}
\end{figure}

To explore the nature of QPTs with different staggered biases, we will
discuss the order parameter and the entanglement entropy in detail. We
extract the data of the order parameter and the entropy as a function of
coupling strength $\alpha $ along $\epsilon =0.1$, $0.2$ for $s=0.3$, and $%
\epsilon =0.5$, $0.7$ for $s=0.7$, and replot in the upper panels of Figs. %
\ref{QPT_s03} and \ref{QPT_s07}, respectively. It is found that the order
parameter (blue line) becomes nonzero continuously for $\epsilon =0.1$ at $%
s=0.3$ and $\epsilon =0.5$ at $s=0.7$, indicating a 2nd-order QPT, while it
suddenly jumps to a finite value for $\epsilon =0.2$ at $s=0.3$ and $%
\epsilon =0.7$ at $s=0.7$, suggesting a 1st-order QPT. By extensive
calculations in a similar way, we can immediately locate a critical point
that splits the whole critical line into the 1st- and 2nd-order critical
lines, as indicated in the upper panels of Fig. \ref{phase_diagram} with red
dots. This is just a QuTP, similar to that observed in the generalized Dicke
model in the staggered biases \cite{puhan}.

The same picture can also be drawn from the entanglement entropy indicated
with red lines in the upper panels of Figs. \ref{QPT_s03} and \ref{QPT_s07}.
For $\epsilon =0.1$ at $s=0.3$ and $\epsilon =0.5$ at $s=0.7$, the entropy
of entanglement $S_{E}$ displays a cusplike behavior, similar to that
observed in the single sub-Ohmic SBM~\cite{Chin,lehur}, thus demonstrating
the 2nd-order QPT. Whereas, for large $\epsilon $, e.g. for $\epsilon =0.2$
at $s=0.3$ and $\epsilon =0.7$ at $s=0.7$, although the entropy still
displays a sharp nonanalyticity at the transition point, it suddenly drops
to a finite value, in contrast to the 2nd-order QPT for small $\epsilon $
where $S_{E}$ falls off gradually on  both sides of the phase transition
point. As shown in the upper left panels of Figs. \ref{QPT_s03} and \ref%
{QPT_s07}, the sudden drop of the entropy occurs simultaneously at the
sudden jump of the order parameter at the same $\epsilon$, thus both
suggesting the 1st-order QPTs.

\begin{figure}[tbp]
\centerline{\includegraphics[scale=0.4]{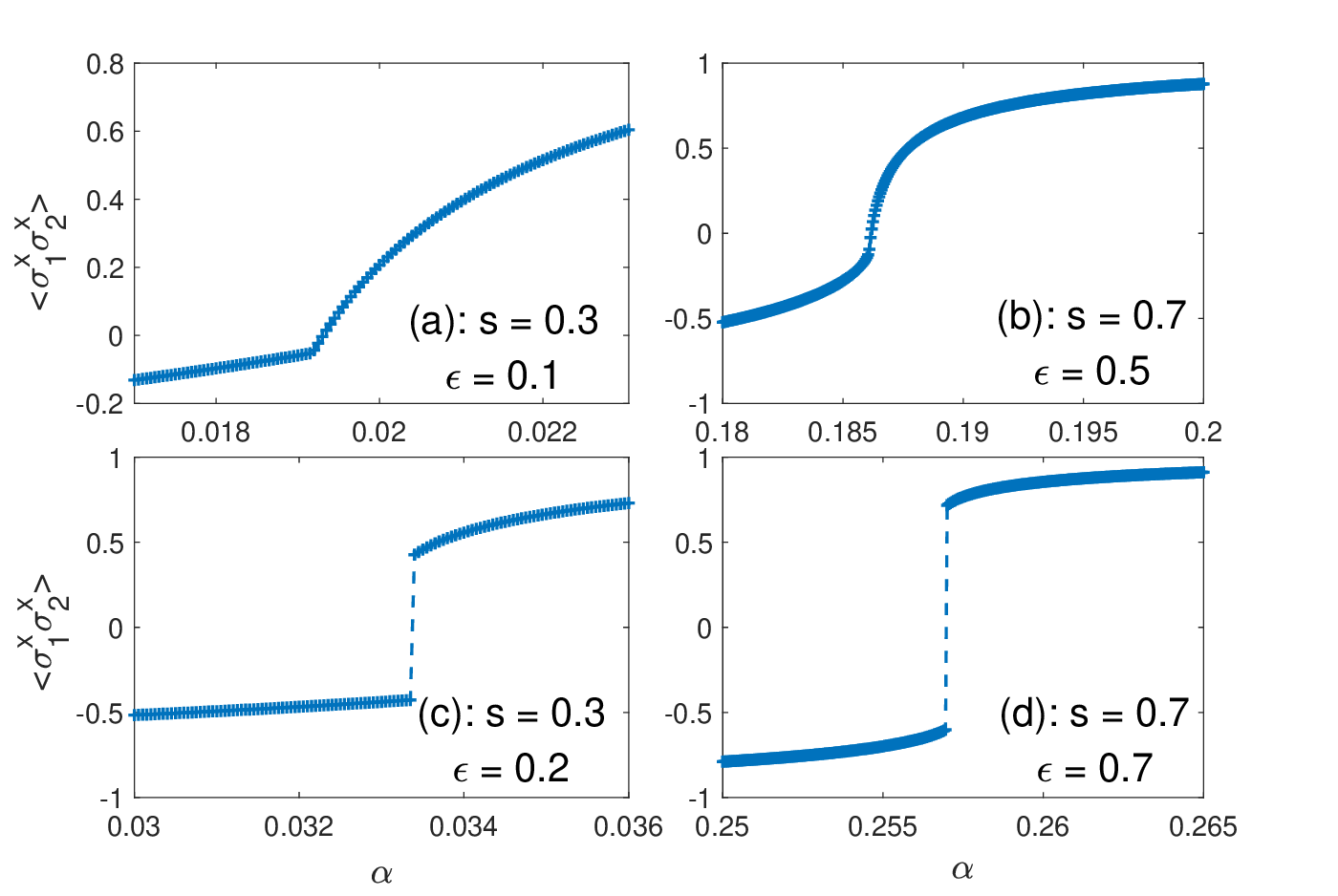}}
\caption{ (Color online) The correlation function $\left\langle \protect%
\sigma _{1}^{x}\protect\sigma _{2}^{x}\right\rangle $ as a function of the
coupling strength {\ for the sub-Ohmic bath} by VMPS approaches at $\protect%
\epsilon =0.1,0.2$ for $s=0.3$ (left) and $\protect\epsilon =0.5,0.7 $ for $%
s=0.7$ (right) . Other parameters: $\Delta =0.1$, $\protect\omega _{c}=1 $, $%
\Lambda =2$, $L=50$, $d_{opt}=12$, $D=20$.}
\label{correlation}
\end{figure}
The 1st-order and 2nd-order QPTs can also be directly discerned by the
first- and second-order derivatives of the ground-state energy with respect
to the coupling parameter $\alpha $. The results at the same model
parameters are presented in {the lower panels of Figs. \ref{QPT_s03} and \ref%
{QPT_s07}. At the two smaller staggered bias (lower left), the 1st-order
derivatives of the energy are continuous around the transitions, while at
the two larger staggered biases (lower right), they are discontinuous at the
critical points, whereas  the 2nd-order derivatives of energy are
discontinuous for smaller $\epsilon$ and diverge for the larger $\epsilon$
at the phase transition points for each bath exponent $s$, respectively. The
observations based on the ground-state energy are obviously in accord with
the original criterion of the 2nd- and 1st-order phase transitions,
justifying again the existence of QuTP in the phase diagram based on the
order parameter and the entropy.}

To provide further evidence of the existence of the QuTP separating the 1st-
and 2nd-order critical lines, we calculate the two spin correlation function
$\left\langle \sigma _{1}^{x}\sigma _{2}^{x}\right\rangle $. The results are
shown in \ Fig. \ref{correlation} for $s=0.3$ (left) and $0.7$ (right), at
small and large biases, which are the same as those in Figs. \ref{QPT_s03}
and \ref{QPT_s07}. It is observed that the $\left\langle \sigma
_{1}^{x}\sigma _{2}^{x}\right\rangle $ is continuous (discontinuous) for
small (large) staggered biases, also demonstrating the 2nd (1st)-order QPTs
at the corresponding bias.

\begin{figure}[tbp]
\centerline{\includegraphics[scale=0.4]{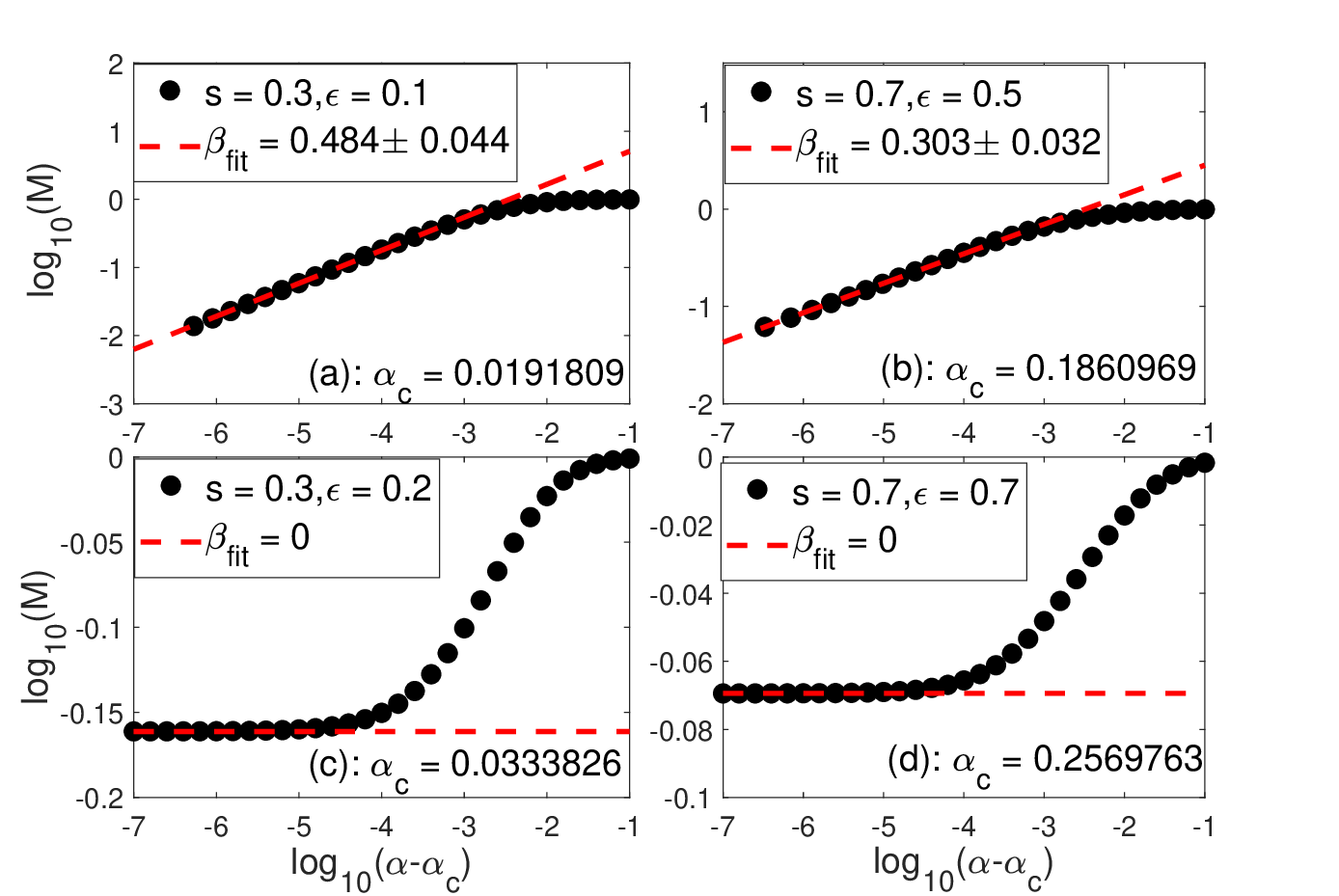}}
\caption{ (Color online) The log-log plot of the magnetization $M $ as a
function of $\protect\alpha-\protect\alpha_c $ at $\protect\epsilon =
0.1,0.2 $ for $s = 0.3$ (left panels) and $\protect\epsilon = 0.5,0.7$ for $%
s = 0.7$ (right panels). The numerical results by VMPS are denoted by black
circles and the power-law fitting curves are denoted by the red dashed
lines, which indicates the 2nd-order QPT takes place in the smaller
staggered biases and gives similar critical behaviors compared to the
standard spin-boson model, while the larger staggered biases induce the
1st-order QPT and vanishing of the critical exponent $\protect\beta$. $%
\Delta =0.1$,$\protect\omega _{c}=1$, $\Lambda=2 $, $L=50$, $d_{opt}=12$,
and $D=20,40$ for $s=0.3,0.7$ respectively.}
\label{exponent}
\end{figure}

Since the staggered biases can drive the 2nd-order QPTs to the 1st-order
ones, can it alter the universality class in the 2nd-order critical lines? \
In order to answer this question, we{\ present the log-log plot of the
magnetization }$M= \left( \left\langle \sigma _{1}^{x}\right\rangle
+\left\langle \sigma _{2}^{x}\right\rangle \right) /2 ${\ as a function of $%
\alpha -\alpha _{c}$ in Fig. \ref{exponent} where the parameters are the
same as those in Fig. \ref{correlation}. The critical exponents $\beta $ can
be determined by fitting power-law behavior, ${M{\propto }\left( {\alpha -{%
\alpha _{c}}}\right) ^{\beta }}$. For two smaller } biases below the QuTP,
as displayed in the upper panels of {Fig. \ref{exponent}, very nice
power-law behavior over three decades is demonstrated for both cases,
yielding }$\beta =0.484$ for ${s=0.3}$ and $\beta =0.303$ for $s=0.7$, which
are very close to those in the single SBM for the same $s$ by the VMPS
approaches ~\cite{guo2012critical}. This is to say, the critical exponents
of the order parameter {\ are not different from those in the single SBM. In
other words, as long as the 2nd-order QPTs occurs in the 2SBM with the
staggered biases, the critical exponent is only the bath dependent, and
remains unchanged with }$\epsilon ${. At the 1st-order critical line in the
large }$\epsilon $ regime, as shown in the low panels of {Fig. \ref{exponent}%
, $\beta =0$, consistent with the 1st-order phase transition nature.}

\subsection{Ohmic bath (s=1)}

It is well known that the single SBM with the Ohmic bath undergoes the
continuous QPTs of KT type ~\cite{Leggett}. In the language of the
quantum-to-classical mapping, $s=1$ corresponds to the low critical
dimension of the long-ranged Ising model~\cite{Luijten}. As shown in the
last section, in the sub-Ohmic 2SBM, the staggered biases can drive the
2nd-order QPT to the 1st-order one. Then what is the effect of these
staggered biases on the KT phase transitions in the Ohmic 2SBM? Could the
staggered biases drive the KT phase transitions to the 2nd-order or/and the
1st-order ones?

\begin{figure}[tph]
\centerline{\includegraphics[scale=0.4]{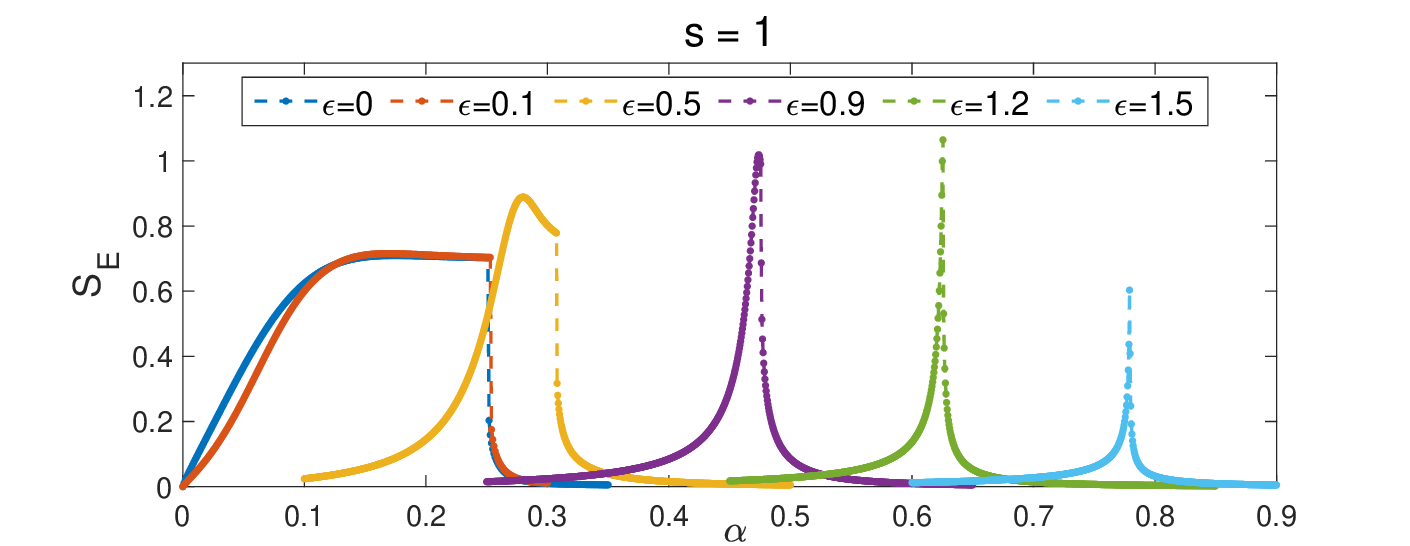}}
\caption{ (Color online) Entanglement entropy $S_{E} $ as a function of $%
\protect\alpha $ in the ground state {\ for the Ohmic bath ($s=1$) } at $%
\protect\epsilon =0,0.1,0.5,0.9,1.2,1.5$ for $s = 1$ by VMPS approach. $%
\Delta =0.1$, $\protect\omega _{c}=1$, $\Lambda =2$, $L=50$, $d_{opt}=12$,
and $D=20$.}
\label{entropy_s1}
\end{figure}
To address these issues, we also extend to study 2SBM in the Ohmic bath with
the staggered biases using the VMPS in this subsection. In the literature,
the entanglement entropy is usually studied in the SBM with the Ohmic bath,
because KT phase transitions are of infinite order, and less observables can
be used to distinguish KT from other kinds of phase transitions. In the KT
phase transition of the single SBM for $s=1$, the entropy increases in the
weak coupling regime, then saturates to a plateau, and drops suddenly at the
KT critical point \cite{lehur_kt}. The sudden drop of the entanglement
entropy signifies the onset of an emergent new phase. In the 2nd-order QPTs,
the entropy falls off gradually in both sides of the critical point \cite%
{Chin,lehur}, displaying different behavior from those in both the KT and
the first-order QPTs.

We calculate the entanglement entropy for several staggered biases from $%
\epsilon =0$ to $1.5$ in Fig. \ref{entropy_s1}. We find that for all values
of $\epsilon $, the entropy suddenly drops at a critical point, exhibiting a
sharp nonanalyticity, and therefore signifying the emergence of a different
phase. The sudden drop of the entropy demonstrates either the 1st-order or
KT phase transitions, thus excluding the 2nd-order QPT.

\begin{figure}[tph]
\centerline{\includegraphics[scale=0.4]{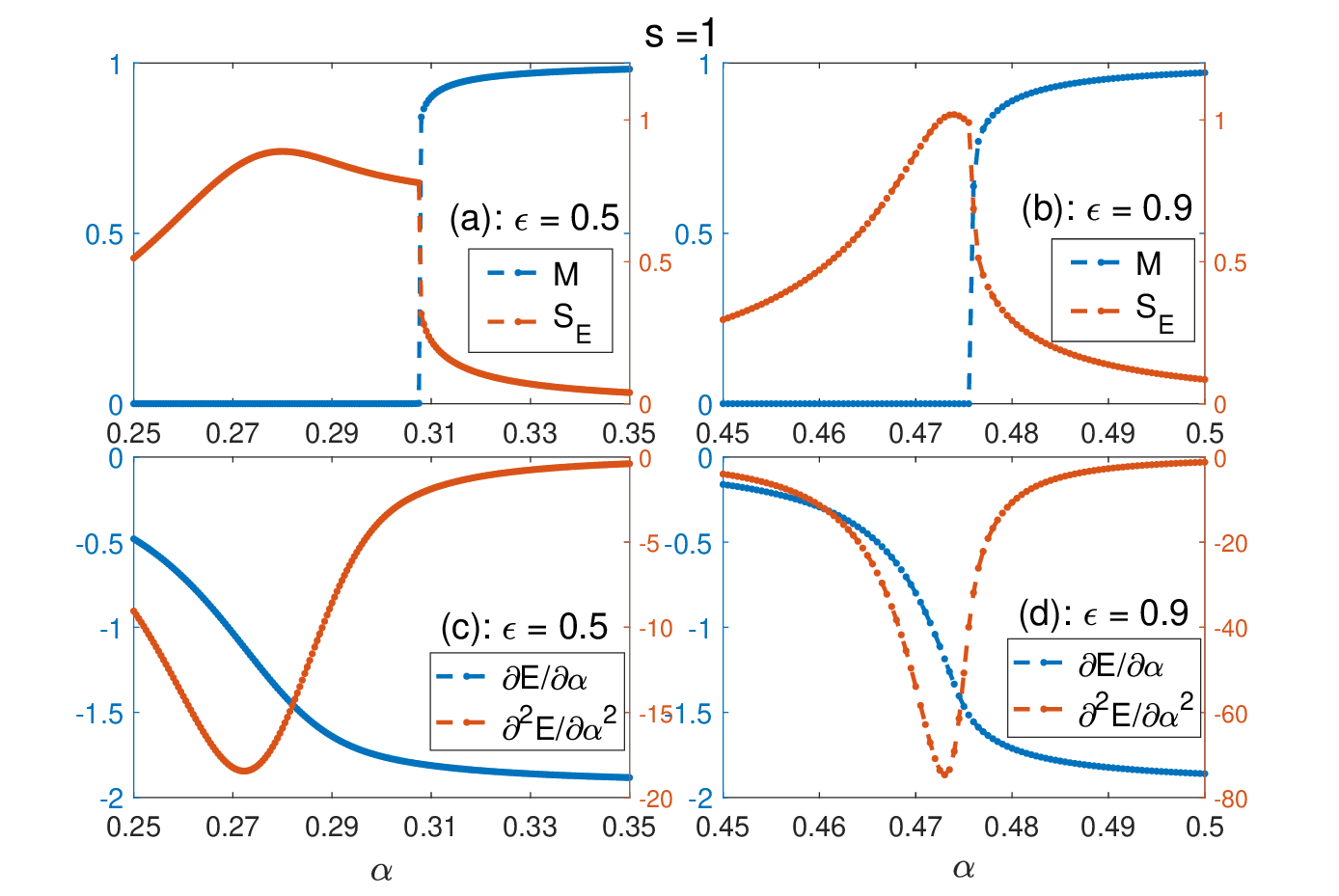}}
\caption{ (Color online) Magnetization $M$, entanglement entropy $S_{E} $
(upper panels), the first- and second-order derivatives of the ground-state
energy (lower panels) {\ as a function of $\protect\alpha $ in the Ohmic
bath ($s=1$) for a weak bias $\protect\epsilon =0.5$ (left panels) and
strong bias $\protect\epsilon =0.9$(right panels), representative of the KT
transition, } by VMPS approach. $\Delta =0.1$, $\protect\omega _{c}=1$, $%
\Lambda =2 $, $L=50$, $d_{opt}=12$, and $D=20$ for $s=1$. }
\label{QPT_s1_small}
\end{figure}

With the increasing staggered biases, the flat plateau gradually changes
into a broad peak, and shrinks considerably at rather large $\epsilon $. To
be more clear, we replot the entropy at two typical staggered biases $%
\epsilon =0.5$ and $0.9$ in the enlarged view in the upper panels of Fig. %
\ref{QPT_s1_small}. Interestingly, at $\epsilon =0.5$, the entropy shows a
broad peak before dropping abruptly at the transition point, different from
that in the single SBM at $s=1$ where the entropy saturates at a wide
coupling range before a sudden drop at the transition point \cite{lehur_kt}.
We argue that the coherence is lost already before the system becomes
localized~ \cite{Lehur2} due to the presence of the staggered biases, so the
flat plateau decays to a broad peak at the finite but small $\epsilon $. At $%
\epsilon =0.9$, the maximum point of the narrow peak is very close to, but
still not at the transition point, {\ in contrast to the sub-Ohmic SBM where
the maximum of entanglement signifies the 2nd-order phase transition}. In
these cases, the phase transition is still of KT type, as will be shown
below.

\begin{figure}[tph]
\centerline{\includegraphics[scale=0.4]{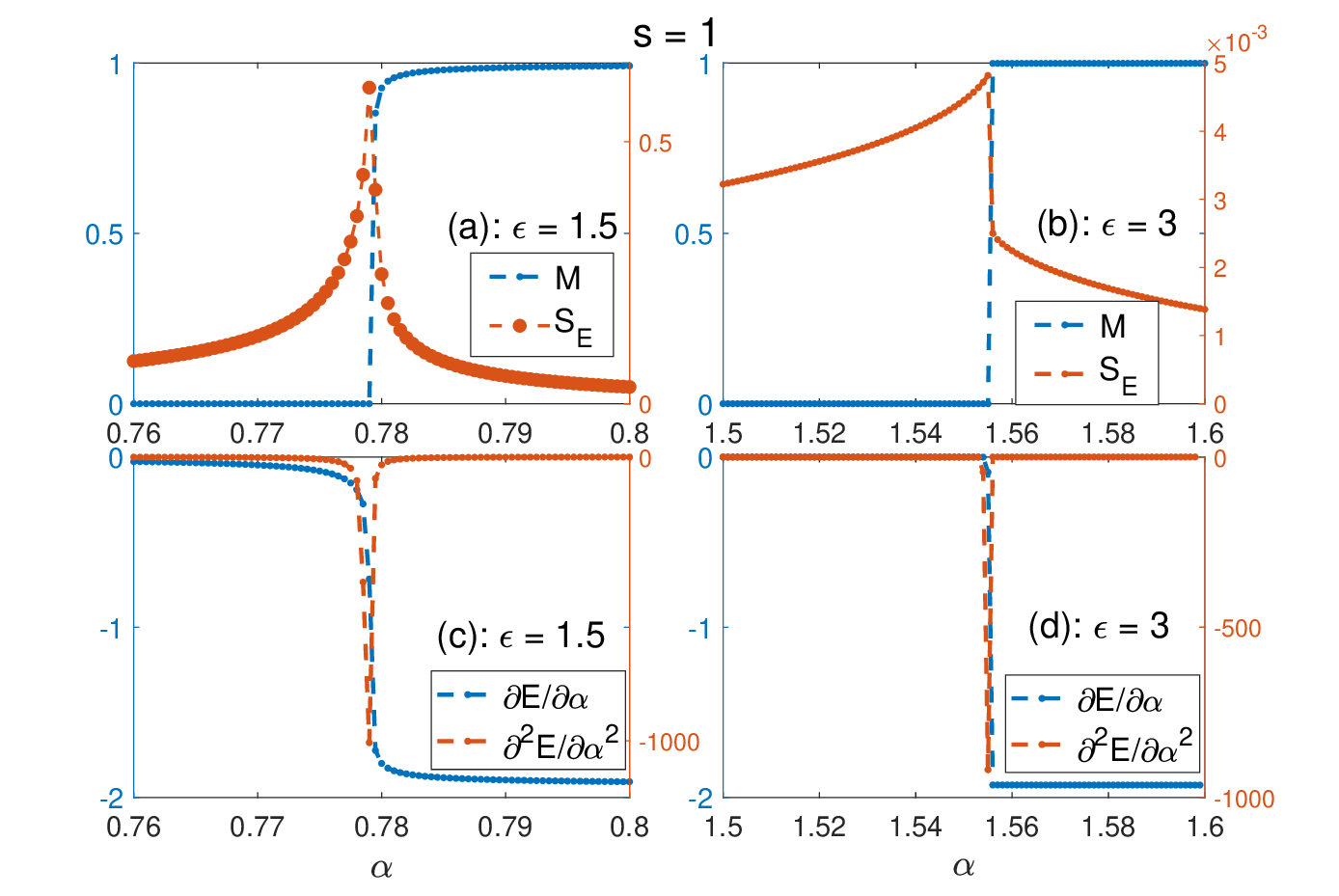}}
\caption{ (Color online) Magnetization $M$, entanglement entropy $S_{E}$
(upper panels), the first- and second-order derivatives of the ground-state
energy (lower panels) as a function of $\protect\alpha $ {\ in the Ohmic
bath ($s=1$) for two strong biases $\protect\epsilon =1.5$ (left panels) and
$\protect\epsilon =3$ (right panels), representative of the first-order
phase transition, } by VMPS approach. $\Delta =0.1$, $\protect\omega _{c}=1$%
, $\Lambda =2$, $L=50$, $d_{opt}=12$, and $D=20$ for $s=1$. }
\label{QPT_s1_large}
\end{figure}

To explore the possible new phase transitions of the 2SBM in the Ohmic bath
driven by the large staggered bias $\epsilon $, we also plot the entropy at
rather large fields in the upper panels of Fig. \ref{QPT_s1_large}. The
entropy  drops abruptly at the transition points for these large staggered
biases. However, for small staggered biases in the upper panels of Fig. \ref%
{QPT_s1_small}, the entropy decreases with $\alpha $ just before the abrupt
drop, contrary to the case at large staggered biases indicated in the upper
panels of Fig. \ref{QPT_s1_large}, where the entropy increases monotonically
before sudden drops. The different behaviors should be originated from
different kinds of phase transitions. The order parameters are then
collected in the upper panels of Figs. \ref{QPT_s1_small} and \ref%
{QPT_s1_large}. However, the order parameter jumps suddenly for all cases at
the transition points, so one could not employ it to discriminate between
the 1st-order and the KT type QPTs.

To show the nature of the phase transition in this model, we thus resort to
the first-order and second-order derivatives of the ground-state energy with
respect to $\alpha $, as demonstrated in the low panels of Figs. \ref%
{QPT_s1_small} and \ref{QPT_s1_large} with the same model parameters. It is
found that for small staggered biases, the first- and second-order
derivatives of the ground-state energy at the transition point are
continuous and do not exhibit any exotic behavior. Even the further high
order derivative would not exhibit any discontinuity at the transition
points, displaying the infinite-order KT phase transition nature. However,
for two larger staggered biases, the first-order derivative drops suddenly,
showing discontinuity at the transition point, which is just the typical
characteristics of the 1st-order phase transition.

So in the Ohmic 2SBM with staggered biases, the KT phase transitions can be
directly driven to the first-order one by increasing the staggered biases.
Therefore the QuTP also exists in this model, which separates the 1st-order
and KT critical lines. It is roughly estimated to be $\epsilon \approx 1.18 $%
.

Finally, combining with the observations in the sub-Ohmic case in the last
subsection, we can reach a conclusion that the staggered biases can drive
the original QPTs to the 1st-order ones in 2SBM with both Ohmic and
sub-Ohmic baths directly, and could not change the universality of
continuous phase transitions including the KT phase transitions. We believe
that this conclusion can be generalized to the finite even number of
dissipative spins in the staggered biases.

The universality in the QuTP in the present model is also a challenging
issue. According to the Landau theory, it should be different from those in
other critical points. But it is difficult to use any numerical approaches
to distinguish this isolated point from others. If the analytical treatment
formulated from the Feynman path-integral representation of the partition
function for the single SBM \cite{Leggett,map,berry,Kirchner} can be
extended to this model, then it may be probable to clarify this issue.

{\ Very interestingly, increasing the staggered biases can make the
transition discontinuous in both the spin-boson model and the Dicke model~%
\cite{puhan}. It appears that there should be a common explanation for the
1st-order transitions in both models. There is possibly a third order term
proportional to a power of the bias in the Ginzburg-Landau effective action
in the Feynman path-integral representation. With increasing biases, the
order parameter jumps from zero to a finite value, and results in the
first-order transition directly. We believe that the topic along this line
deserves  further careful study.}

\section{Conclusion}

We have found rich quantum phase transitions in the 2SBM with both the
sub-Ohmic and the Ohmic baths in the staggered biases by the VMPS approach.
The phase diagram has been composed in terms of the coupling strength and
the bias magnitude. For the sub-Ohmic bath, we find that the 2nd-order
critical lines meet the first-order ones at the QuTP. For the Ohmic bath, we
observe that { the KT phase transitions can be driven } directly
to the 1st-order phase transitions. \ For all cases, if the 1st-order phase
transition does not emerge, \ the universality of the phase transition could
not be changed by the applied staggered biases.

The recent superconducting circuit QED system has allowed for the SBM in
both the Ohmic and the sub-Ohmic bath~\cite{Forn2,Yamamoto,exp_SB,Nori},
thus the 2SBM is experimentally feasible. Unlike the conventional cavity QED
systems, the static bias of the qubit present in the circuit QED systems is
ubiquitous, and can be easily introduced and manipulated by an externally
applied magnetic flux \cite{Niemczyk,Yoshihara}, which provides an
additional dimension to exhibit the rich QPTs. Generalized Dicke models
without the nonlinear Stark coupling undergo the 2nd-order QPT in the
thermodynamic limit (i.e. infinite atomic number $N\rightarrow \infty $)
\cite{puhan}, while the finite-component QPT requires to implement an
extremely large detuning (i.e. infinite frequency ratio $\Delta /\omega
\rightarrow \infty $) ~ \cite{QRM,Felicetti,GFZhang}, so the present
considered phase transition in the 2SBM in the staggered biases might be
easier to realize experimentally. We believe that the 2SBM would become a
potential platform to test the rich quantum criticality and the QuTP.

\textbf{ACKNOWLEDGEMENTS} We acknowledge useful discussions with Stefan
Kirchner. This work is supported by the National Key Research and
Development Program of China under Grant No. 2017YFA0303002 and the National
Science Foundation of China under Grant No. 11834005.

$^{\ast }$ Corresponding author. Email: qhchen@zju.edu.cn


\end{document}